\documentclass[twocolumn,showpacs,preprintnumbers,amsmath,amssymb]{revtex4}

\usepackage{graphicx}
\usepackage{dcolumn}
\usepackage{bm}

\begin{document}
\title {High-order inertial phase shifts for time-domain atom interferometers}
\author{Kai Bongs, Romain Launay, and Mark A. Kasevich}
\affiliation{Physics Department, Yale University, New Haven CT,
06520}

\date{\today}

\begin{abstract}
High-order inertial phase shifts are calculated for time-domain
atom interferometers.  We obtain closed-form analytic expressions
for these shifts in accelerometer, gyroscope, optical clock and
photon recoil measurement configurations. Our analysis includes
Coriolis, centrifugal, gravitational, and gravity gradient-induced
forces. We identify new shifts which arise at levels relevant to
current and planned experiments.
\end{abstract}


\maketitle Atom interferometric measurements have growing
applications in basic and applied science.  For example, atom
interferometric techniques have been recently used to measure
rotations \cite{Gustavson_2000}, gravity gradients
\cite{Snadden_1998}, $\hbar/m_{\text{Cs}}$ \cite{Weiss_x993} and
accelerations \cite{Peters_x999} with unprecedented precision.
Future applications range from tests of General Relativity to the
development of next generation inertial navigation systems. The
accuracy of atom interferometric instruments, however, hinges on
the accuracy of the theory used to connect the measured
interferometric phase shift to the physically relevant quantities.

In this paper, we develop analytic expressions for the response of
commonly used atom interferometer measurement configurations to
experimentally relevant combinations of rotation, gravity and
gravity gradient-induced forces.  We identify new classes of
high-order phase shifts which are observable in current
experiments, and which seem to be of vital importance for proposed
future experiments.  In particular, we show below that current
acceleration measurements need to be corrected at the part per
billion (ppb) level for high-order rotation effects, that
measurements of the photon recoil need a 10 ppb correction for
gravity gradients, and that next generation optical time standards
will need gravity gradient corrections to achieve fractional
frequency accuracies of $\delta \nu/ \nu$  below $10^{-17}$.
Related theoretical work in this area has treated simple
(one-dimensional) models \cite{Peters_x999,gg_1998} or provided
general frameworks \cite{Borde_2001}.

\begin{figure}
\includegraphics[width=\columnwidth]{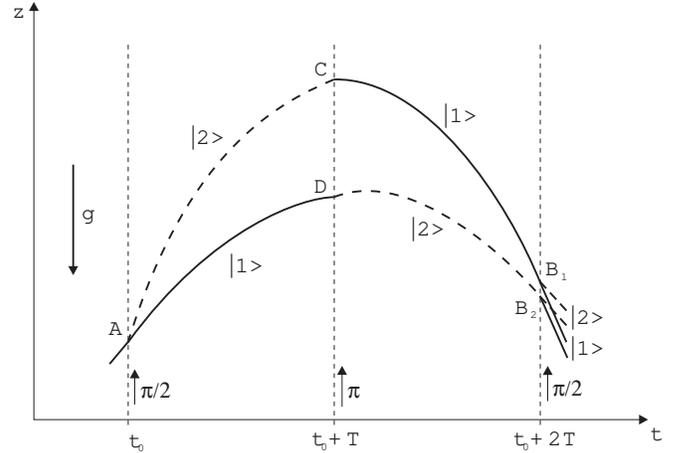}
\caption{\label{fig_1} Schematic illustration of the
interferometer accelerometer geometry. For this realization,
$\Delta\phi_{\text{prop}}=({S_{\text{cl,ACB$_1$}}-S_{\text{cl,ADB$_2$}}})/{\hbar}$
with $S_{\text{cl,ACB$_1$}}$ and $S_{\text{cl,ADB$_2$}}$
representing the classical action along the interfering paths
ACB$_1$ and ADB$_2$.}
\end{figure}

For simplicity we confine our discussion to the case of
time-domain light-pulse atom interferometers \cite{Kasevich_1991}.
However, with minor modification our results can be applied to
other de Broglie wave interferometry approaches
\cite{Berman_1998}.  For light-pulse atom interferometers,
coherent division, redirection and recombination of atomic
wavepackets is accomplished via momentum exchange with an external
driving laser field \cite{Borde_x989}.  Complete descriptions of
experimental realizations based on this principle are given in
Ref. \cite{Berman_1998}.  In brief, an optical pulse of area
$\pi/2$, coupling two stable internal atomic states $|1\rangle $
and $|2\rangle $, serves to coherently divide an incident atomic
wavepacket:  an atom initially in state $|1\rangle $ is driven
into a coherent superposition of internal states $|1\rangle $ and
$|2\rangle $, with the momenta of the wavepackets associated with
these states differing by the momentum of the photon used to drive
the transition.  Pulses of area $\pi $ induce stimulated
transitions which exchange the internal states associated with the
atomic wavepackets, while changing their momenta by the photon
recoil momentum.  Atom interferometers are realized from sequences
of these pulses.  For example, $\pi/2 - \pi - \pi/2$ pulse
sequences have been used to build atomic interferometers analogous
to optical Mach-Zehnder interferometers, as illustrated in Fig.
\ref{fig_1} \cite{Kasevich_1991}.

The total phase shift between interfering paths,
$\Delta\phi_{\text{total}}$, can be broken into three
contributions:
\begin{equation} \nonumber
\Delta\phi_{\text{total}} =
\Delta\phi_{\text{prop}}+\Delta\phi_{\text{laser}}+\Delta\phi_{\text{sep}},
\end{equation}
where $\Delta\phi_{\text{prop}}$ is the phase shift due to
wavepacket propagation between the interrogating optical pulses,
$\Delta\phi_{\text{laser}}$ is the shift acquired during the
laser-atom interactions used to manipulate the atomic wavepackets,
and $\Delta\phi_{\text{sep}}$ is the shift due to the (possible)
final spatial separation of the interfering wavepackets at the
interferometer output port \cite{Storey_1994}.  Our approach
neglects terms originating in the spatial extent of the
wavepackets \cite{Borde_2001} and invokes the short-pulse (high
Rabi frequency) limit for the optical interactions in order to
clarify contributions arising from inertial forces.  We briefly
summarize expressions for $\Delta\phi_{\text{prop}}$,
$\Delta\phi_{\text{laser}}$ and $\Delta\phi_{\text{sep}}$ below.

The $\Delta\phi_{\text{prop}}$ term is obtained using the Feynman
path integral approach \cite{Feynman_Hibbs}, which involves
calculation of the difference between the classical actions
associated with the interfering wavepacket trajectories.  This is
illustrated in Fig. 1 for the case of the accelerometer
implementation.  The actions $S$ are obtained by integrating the
Lagrangian $L$ over the classical trajectories $\Gamma$ associated
with the mean positions of each wavepacket, {\it e.g.} $ S =
\int_{\Gamma} L(\mathbf{r}(t),\mathbf{v}(t))dt$, with the path
$\Gamma$ determined by the classical trajectory $(\mathbf{r}(t),
\mathbf{v}(t))$.  This approach is formally correct when $L$ is at
most second order in position $\mathbf{r}(t)$ and velocity
$\mathbf{v}(t)$ (see Ref. \cite{Heller_1975}).  We find the
classical trajectories $\Gamma$ through integration of the
Euler-Lagrange equations for $L$.

The phase difference resulting from laser interactions
$\Delta\phi_{\text{laser}}$ is discussed in detail in Ref.
\cite{Berman_1998}. This shift is obtained from solution of the
Schr\"odinger equation for a resonantly driven two-level quantum
system. In the limit of impulse excitations, this solution reduces
to the following rules for the evolution of the probability
amplitudes of states $|1 \rangle$ and $|2\rangle$:
\begin{equation}
  |1\rangle \rightarrow i \exp[i{\bf k} \cdot \mathbf{r}(t_p)] |2\rangle; \; \; \; \;
  |2\rangle \rightarrow i \exp[-i{\bf k} \cdot
  \mathbf{r}(t_p)] |1\rangle.
\end{equation}
Here $\bf{k}$ is the propagation vector associated with the
coupling laser field(s) and $\mathbf{r}(t_p)$ is the mean position
of the atomic wavepacket at the interaction time $t_p$. Time
dependent terms associated with the frequency of the driving
fields have been suppressed (as they cancel for the time symmetric
pulse sequences considered here), and we assume the initial laser
phase to be constant during the interferometer sequence. As an
example, for the pulse sequence of Fig. \ref{fig_1}, the resulting
laser-induced phase difference for an ensemble initially prepared
in state $|1\rangle $ and detected in state $|1\rangle$ is:
\begin{equation}
\Delta\phi_{\text{laser}} = \bf k \cdot
(\mathbf{r}_{\text{A}}-\mathbf{r}_{\text{C}}-\mathbf{r}_{\text{D}}
+ \mathbf{r}_{\text{B}_2}).
\end{equation}

The shift $\Delta \phi_{\text{sep}}$ arises when the classical
positions of the two interfering trajectories do not coincide at
the exit beamsplitter (as illustrated in Fig. \ref{fig_1}).  This
shift is adequately approximated by
\begin{equation}
\Delta\phi_{\text{sep}} = \mathbf{p}\cdot \Delta \mathbf{r}/\hbar,
\end{equation}
where $\mathbf{p}$ is the mean momentum of the wavepackets in a
given output port, and $\Delta \bf{r} $ is the spatial separation
between the centers of each wavepacket at the time of the last
optical pulse \cite{momentum}. In certain special cases -- for
example, that of uniform gravity -- this contribution is zero.

\begin{figure}
\includegraphics[width=0.7\columnwidth]{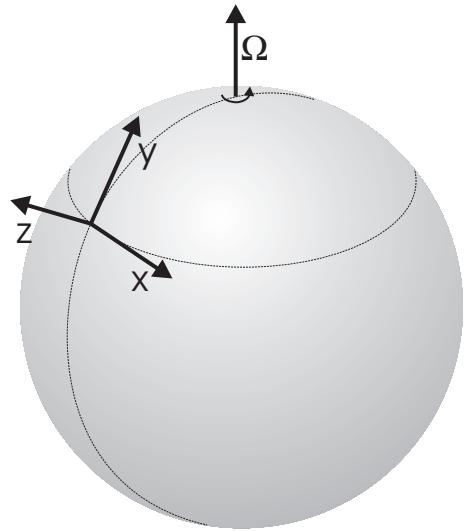}
\caption{\label{fig_2} Reference frame for the calculations. The
z-axis is chosen to point away from the Earth center and the
x-axis is tangent to a latitude circle pointing east.}
\end{figure}

The above phase shifts were evaluated in a geocentric reference
frame fixed to the surface of the Earth, as shown in Fig.
\ref{fig_2}. In this frame, the Lagrangian can be written as:
\begin{eqnarray}\label{full_L} L(\mathbf{r},\mathbf{v}) &=&  m\left( \frac{\mathbf{v}^2}{2}
+ \mathbf{g} \cdot \mathbf{r} + \frac{1}{2}r_i T_{ij} r_j
 + \mathbf{\Omega} \cdot \left( (\mathbf{r} + \mathbf{R} )\times \mathbf{v} \right) \right. \nonumber \\
 & & \left. + \frac{1}{2}\left( \mathbf{\Omega} \times (\mathbf{r} + \mathbf{R}) \right)^2
 \right),
\end{eqnarray}
where $m$ is the atomic mass, $\mathbf{R}$ is the displacement
from the center of the Earth, $\mathbf{\Omega}$ the Earth rotation
rate, $\mathbf{g}$ the acceleration due to gravity and $T_{ij}$ is
the gravity gradient tensor.

Analytic solutions were obtained in a perturbative approach
\cite{Storey_1994}, where the action was evaluated for the full
Lagrangian $L$ over trajectories determined by the Lagrangian
$\tilde L$ which neglected gravity gradients, {it e.g.}:
\begin{eqnarray}
\tilde L(\bf r,\bf v) &=& m\left( \frac{\bf{v}^2}{2} + \bf g \cdot
\bf r + \bf \Omega \cdot \left( (\bf r + \bf R )\times \bf v \right) \right.  \nonumber \\
& & \left.      + \frac{1}{2}\left( \bf \Omega \times (\bf r + \bf
R ) \right)^2 \right).
\end{eqnarray}
This approximation was validated numerically as discussed below.

We evaluated accelerometer \cite{Kasevich_1991, Peters_x999},
gyroscope \cite{Gustavson_2000}, photon recoil \cite{Weiss_x993},
and optical clock interferometer \cite{Borde_x989} configurations,
as summarized in Tables \ref{table_1}-\ref{table_5}.  Each
configuration is discussed briefly below.  We tabulate only the
most significant terms in Taylor expansions of the total phase
shifts. In order to give a quantitative measure for the different
phase contributions, the tabulated expressions were evaluated for
parameters corresponding to current experiments.  In particular,
$m=2.21\cdot 10^{-25}\, $kg (Cs atomic mass),
$\lambda_{\text{eff}} =426 \, $nm (effective wavelength for
two-photon Raman transitions), latitude $41^{\circ }$, Earth
radius $R=6.72 \cdot 10^6 \, $m, gravity $g_z=-9.8 $ m/sec$^2$ and
gravity gradient $2T_{xx}=2T_{yy}=-T_{zz}=2g_z/R$ with all other
gravity gradient terms being zero (for a spherically symmetric
Earth).  The coordinate system origin is taken as the mean
wavepacket position at the time of the first optical pulse, and
the phase shift is determined just after the final pulse.

\medskip \noindent {\it Gravimeter.} Phase shifts for an
atomic fountain gravimeter/accelerometer, based on a $\pi/2 - \pi
- \pi/2$ sequence of vertically propagating optical pulses, are
presented in Table \ref{table_1}.  For the numerical estimates we
used a time $T=0.4$ s between the pulses (corresponding to a
fountain height of $78$ cm) and a vertical launch velocity
$v_z=-g_zT$.  Note that in a satellite microgravity experiment the
first two terms would cancel, such that the gravity gradient
contribution would dominate.  Recent terrestrial experiments have
demonstrated the capability to resolve phase shifts at or below
the $10^{-9}$ g level \cite{Peters_x999}.

\begin{table}
\caption{\label{table_1} Gravimeter phase shifts.}

\begin{ruledtabular}
\begin{tabular}{c c c}
   Term                                     & Phase (rad)                & Relative phase \\ \hline
  $k_zT^2g_z$                               & $-2.32\cdot 10^7$    & $1.0$ \\
  $k_zT^2\Omega_y^2R$                       & $4.44\cdot 10^4$     & $1.9\cdot 10^{-3}$ \\
  $k_zT^3v_zT_{zz} $                        & $1.08\cdot 10^{1}$   & $4.7\cdot 10^{-7}$ \\
  $\frac{7}{12}k_zT^4g_zT_{zz}$             & $-6.32$              & $2.7\cdot 10^{-7}$ \\
  $-3k_zT^3v_z\Omega_y^2$                   & $-3.11\cdot 10^{-2}$ & $1.3\cdot 10^{-9}$ \\
  $-\frac{7}{4}k_zT^4g_z\Omega_y^2$         & $1.81\cdot 10^{-2}$  & $7.8\cdot 10^{-10}$ \\
  $\frac{7}{12}k_zT^4T_{zz}\Omega_y^2R$     & $1.21\cdot 10^{-2}$  & $5.2\cdot 10^{-10}$ \\
  $\frac{\hbar}{2m}k_z^2T^3T_{zz}$          & $9.71\cdot 10^{-3}$  & $4.2\cdot 10^{-10}$ \\
  $-\frac{7}{4}k_zT^4\Omega_y^4R$           & $-3.47\cdot 10^{-5}$  & $1.5\cdot 10^{-12}$ \\
  $-\frac{3\hbar}{2m}k_z^2T^3\Omega_y^2$    & $-2.79\cdot 10^{-5}$  & $1.2\cdot 10^{-12}$ \\
  $-\frac{7}{4}k_zT^4\Omega_y^2\Omega_z^2R$ & $-2.62\cdot 10^{-5}$  & $1.1\cdot 10^{-12}$ \\
\end{tabular}
\end{ruledtabular}
\end{table}

\medskip \noindent {\it Optical clock.} The optical Ramsey pulse
sequence, $\pi/2 \uparrow - \pi/2 \uparrow - \pi/2 \downarrow -
\pi/2 \downarrow $ (arrows indicate propagation directions of the
interrogating pulses) is presented in Table \ref{table_3}. Results
are shown for pulses propagating along a vertical axis, with
$T=0.4\, $s between the first and second and the third and fourth
pulse, infinitesimal time between the second and third pulse and
initial vertical velocity $v_z=-g_zT$. Atomic clocks based on this
sequence offer the prospect of pushing the accuracy of the
definition of the second below the $10^{-18}$ level and are a
longstanding goal in frequency metrology.  For an operational time
standard, terms linear in $k_z$ can be suppressed by pulse
reversal techniques \cite{Trebst_2001} ($k_z \rightarrow -k_z$,
for example), leaving terms quadratic in $k_z$ as possible
systematic shifts. The largest of these tabulated is the
well-known recoil-shift. The smaller term, which is linear in
$T_{zz}$, depends on the location of the measurement, and
represents a possible systematic offset at the $\delta \nu / \nu
\sim 10^{-17}$ level.  This term exists for both horizontal and
vertical interrogation geometries.

\begin{table}
\caption{\label{table_3} Phase terms for an optical clock with
light pulses parallel to gravity.}

\begin{ruledtabular}
\noindent \begin{tabular}{c c c}
   Term                                  & Phase (rad)                & Relative phase \\ \hline
  $k_zT^2g_z$                            & $-2.32\cdot 10^7$    & $1.0$ \\
  $k_zT^2\Omega_y^2R$                    & $4.44\cdot 10^4$     & $1.9\cdot 10^{-3}$ \\
  $-k_z^2T\frac{\hbar }{m}$              & $-4.16\cdot 10^4$    & $1.8\cdot 10^{-3}$\\
  $k_zT^3v_zT_{zz}$                      & $1.08\cdot 10^1$     & $4.7\cdot 10^{-7}$\\
  $\frac{7}{12}k_zT^4g_zT_{zz}$          & $-6.32$              & $2.7\cdot 10^{-7}$\\
  $-3k_zT^3v_z\Omega_y^2$                & $-3.11\cdot 10^{-2}$ & $1.3\cdot 10^{-9}$\\
  $-\frac{7}{4}g_zk_zT^4\Omega_y^2$      & $1.81\cdot 10^{-2}$  & $7.8\cdot 10^{-10}$\\
  $\frac{7}{12}k_zT^4RT_{zz}\Omega_y^2$  & $1.21\cdot 10^{-2}$  & $5.2 \cdot 10^{-10}$ \\
  $\frac{\hbar}{3m}k_z^2T^3T_{zz}$       & $6.48\cdot 10^{-3}$  & $2.7\cdot 10^{-10}$\\
\end{tabular}
\end{ruledtabular}
\end{table}

\medskip \noindent {\it Photon recoil measurement.} Chu and
coworkers have shown that a modified form of the optical Ramsey
method can be used for precise determination of the quantity
$\hbar/m_{\text{Cs}}$ \cite{Weiss_x993}. The modification involves
insertion of a series of $N-1$ $\pi $ pulses, of alternating
propagation direction, between the second and third $\pi/2$
pulses, and has the effect of enhancing the photon recoil phase
shift terms.  Following Refs. \cite{Weiss_x993, Chu_2001}, we
calculate the phase difference between the two possible closed
interferometer branches (see Table \ref{table_4}). The phase terms
are evaluated using the following parameters (chosen to correspond
to the experiment in Ref. \cite{Chu_2001}): $T=0.13\, $s for the
time between the first pair as well as the second pair of $\pi/2$
pulses, $T_{\text{rec}}=$ (1/3000) s between the second $\pi/2$
pulse and the first $\pi$ pulse, $T_{\text{rec}}=$ (1/3000) s
between each of the subsequent $N-1$ $\pi $ pulses, and $N-1 = $
30 $\pi$ pulses. The second correction term due to gravity
gradients is at the anticipated level of precision achieved in
recent $\hbar/m_{\text{Cs}}$ measurements.

\begin{table}
\caption{\label{table_4} Phase terms for a photon recoil
measurement.}

\begin{ruledtabular}
\begin{tabular}{c c c}
  Term                                                     & Phase (rad)                  & Relative phase \\ \hline
  $\frac{2N\hbar }{m}k_z^2T$                               & $8.39 \cdot 10^5$       & $1.0$ \\
  $\frac{N\hbar }{3m}k_z^2T^3T_{zz}$                       & $6.89 \cdot 10^{-3}$    & $8.2 \cdot 10^{-9}$ \\
  $\frac{N^2\hbar }{2m}k_z^2T^2T_{\text rec}T_{zz}$          & $8.22 \cdot 10^{-4}$    & $9.8 \cdot 10^{-10}$ \\
  $\frac{(2N^3+N)\hbar }{6m}k_z^2TT_{\text rec}^2T_{zz}$     & $4.36 \cdot 10^{-5}$    & $5.2 \cdot 10^{-11}$ \\
\end{tabular}
\end{ruledtabular}
\end{table}

\medskip \noindent {\it Gyroscope.} A $\pi/2 - \pi  -
\pi/2$ sequence with optical propagation vectors nominally
perpendicular to the mean atomic velocity can be viewed as a
Sagnac-type rotation sensor \cite{Berman_1998}.  In Table
\ref{table_5} we estimate the phase shifts for a time-domain
interferometer with parameters which correspond to the precision
gyroscope of Ref. \cite{Gustavson_2000}. In particular, the light
pulses are chosen to propagate horizontally in the west-east
direction, with the atomic velocity $v_y=290\, $m/s in the
north-south direction and $T=1/290\, $s time between pulses
(corresponding to a $1\, $m spatial separation of the laser
interaction regions).  The largest correction to the well known
Sagnac shift (the leading term) is near the resolution limit of
current instruments, and is compensated by atomic beam reversal
techniques.

\begin{table}
\caption{\label{table_5} Phase terms for a Sagnac rotation
sensor.}

\begin{ruledtabular}
\noindent \begin{tabular}{c c c}
  Term                                            & Phase (rad)                  & Relative phase \\ \hline
  $2k_xT^2\Omega_zv_y$                            & $4.69$                 & $1.0$  \\
  $-2k_xT^3\Omega_yg_z$                           & $6.28 \cdot 10^{-4}$   & $1.3 \cdot 10^{-4}$ \\
  $-2k_xT^3\Omega_y^3R$                           & $-1.20 \cdot 10^{-6}$  & $2.6 \cdot 10^{-7}$ \\
  $-2k_xT^3\Omega_y\Omega_z^2R$                   & $-9.09 \cdot 10^{-7}$  & $1.9 \cdot 10^{-7}$ \\
  $\frac{\hbar }{2m}k_x^2T^3T_{xx}$               & $3.11 \cdot 10^{-9}$   & $6.6 \cdot 10^{-10}$ \\
\end{tabular}
\end{ruledtabular}
\end{table}

\medskip It is interesting to compare the above results with those
from a perturbative treatment of gravity gradients {\it and}
rotations, {\it i.e.} using the Lagrangian $\tilde{\tilde L} =
m\left( {\mathbf{v}^2}/{2} + \bf g \cdot \bf r \right)$ to
determine the classical trajectories, but evaluating the action
with respect to the full Lagrangian $L$ above.  As an example,
Table \ref{table_2} shows these terms for the gravimeter
configuration. Comparison with Table \ref{table_1} indicates that
the phase shift error associated with this commonly used
approximation \cite{Storey_1994} is $\sim 5 \cdot 10^{-3}$ rad or
$\sim 2 \cdot 10^{-10}$ g, which might be resolved by current
experiments.  In contrast, using $\tilde L$ to estimate the
classical trajectories, we obtain agreement at the $\sim$ 1
$\mu$rad level between our analytic calculations and numeric
calculations which use the full Lagrangian $L$ to determine the
classical trajectories.  $\mu$rad level agreement is also obtained
for the optical clock (Table \ref{table_3}) and photon recoil
(Table \ref{table_4}) configurations, while 200 picorad agreement
is obtained for the gyroscope configuration (Table \ref{table_5}).

\begin{table}
\caption{\label{table_2} Phase terms for the accelerometer
sequence derived treating rotations and gravity gradients as
perturbations.}
\begin{ruledtabular}
\begin{tabular}{c c c}
   Term                                    & Phase (rad)                & Relative phase \\ \hline
  $k_zT^2g_z$                              & $-2.32\cdot 10^7$    & $1.0$ \\
  $k_zT^2\Omega_y^2R$                      & $4.44\cdot 10^4$     & $1.9\cdot 10^{-3}$ \\
  $k_zT^3v_zT_{zz} $                       & $1.08\cdot 10^{1}$   & $4.7\cdot 10^{-7}$ \\
  $\frac{7}{12}k_zT^4g_zT_{zz}$            & $-6.32$              & $2.7\cdot 10^{-7}$ \\
  $k_zT^3v_z\Omega_y^2$                    & $1.04\cdot 10^{-2}$  & $4.5\cdot 10^{-10}$ \\
  $\frac{\hbar}{2m}k_z^2T^3T_{zz}$         & $9.71\cdot 10^{-3}$  & $4.2\cdot 10^{-10}$ \\
  $\frac{7}{12}k_zT^4g_z\Omega_y^2$        & $-6.05\cdot 10^{-3}$ & $2.6\cdot 10^{-10}$ \\
\end{tabular}
\end{ruledtabular}
\end{table}

In conclusion, we have analyzed accelerometer, gyroscope, photon
recoil and optical clock interferometer configurations.  We have
identified new phase shift terms which have their origin in
cross-couplings between rotation, acceleration and gravity
gradient perturbations on wavepacket motion.

This work was supported by the Office of Naval Research, Army
Research Office, National Science Foundation and NASA. K. Bongs
thanks the DFG (Deutsche Forschungsgemeinschaft) for financial
support. We thank Neelima Sehgal for technical assistance and
acknowledge valuable conversations with C. Bord\'e and S. Chu.


\begin{thebibliography}{99}

\bibitem{Gustavson_2000}  T.L. Gustavson, A. Landragin and
M.A. Kasevich, Class. and Quant. Grav. {\bf 17}, 2385-2398 (2000);
T.L. Gustavson, P. Bouyer and M.A. Kasevich, Phys. Rev. Lett. {\bf
78}, 2046-9 (1997).

\bibitem{Snadden_1998}  M.J. Snadden, J.M. McGuirk, P.
Bouyer, K.G. Haritos and M.A. Kasevich, Phys. Rev. Lett. {\bf 81},
971-4 (1998); J.M. McGuirk, G.T. Foster, J.B. Fixler, M.J. Snadden
and M.A. Kasevich, Phys. Rev. A {\bf 65}, 033608/1-14 (2002).

\bibitem{Weiss_x993}  D.S. Weiss, B.C. Young and S. Chu,
Phys. Rev. Lett. {\bf 70}, 2706-9 (1993); D.S. Weiss, B.C. Young
and S. Chu, Applied Physics B {\bf 59}, 217-56 (1994).

\bibitem{Peters_x999}  A. Peters, C. Keng Yeow and S. Chu,
Nature {\bf 400}, 849-52 (1999); A. Peters, K.Y. Chung and S. Chu,
Metrologia {\bf 38}, 25-61 (2001).

\bibitem{gg_1998}  P. Wolf and Ph. Tourrenc, Phys. Lett. A
{\bf 251}, 241-246 (1999).

\bibitem{Borde_2001}  C.J. Bord\'e, Comptes Rendus IV
Physique Astrophysique {\bf 2}, 509-530 (2001).

\bibitem{Kasevich_1991}  M. Kasevich and S. Chu, Phys. Rev. Lett.
{\bf 67}, 181-4 (1991); M. Kasevich and S. Chu, App. Phys.
{\bf B54}, 321-32 (1992).

\bibitem{Berman_1998}  P. Berman, {\it Atom Interferometry}
(Academic Press, New York, 1996).

\bibitem{Borde_x989}  C.J. Bord\'e, Physics Letters A {\bf
140}, 10-12 (1989).

\bibitem{Storey_1994} See, for example, P. Storey and C. Cohen-Tannoudji,
Journal de Physique II {\bf 4}, 1999 (1994).

\bibitem{Feynman_Hibbs} R.P. Feynman and A.R. Hibbs,
{\it Quantum Mechanics and Path Integrals} (McGraw-Hill, New York, 1965).

\bibitem{Heller_1975} E. J. Heller, J. Chem. Phys. {\bf 62},
1544 (1975).

\bibitem{momentum} Note that the momentum in the excited state output port differs
from the momentum in the ground state output port by $\hbar \bf
k$. The resulting phase difference between the output ports is
just compensated by changing the $\bf k\cdot \bf r_{\text
B_2}$-term for the ground state output to $\bf k\cdot r_{\text
B_1}$ for the excited state output in the laser phase given above.

\bibitem{Trebst_2001}  T. Trebst, T. Binnewies, J. Helmcke
and F. Riehle, IEEE Trans. Inst. and Meas. {\bf 50}, 535-538 (2001).

\bibitem{Chu_2001} J. Hensley, A. Wicht, B. Young, and S. Chu,
Proc. 17th Int. Conf. Atomic Physics (eds E. Arimondo, P. De
Natale, and M. Inguscio), 43–57 (Am. Inst. Phys., New York, 2000);
A. Wicht, J. Hensley, E. Sarjlic, and S. Chu, Proc. 6th Int'l FSM
Symposium (ed. T. Gill), in press (World Scientific, Singapore,
2002).

\end{thebibliography}
\end{document}